\documentclass[10pt,twocolumn]{IEEEtran}
\usepackage{amssymb}
\usepackage{bbding}
\usepackage{graphicx}
\usepackage{amsmath}
\usepackage{latexsym}
\usepackage[table]{xcolor}
\usepackage{booktabs}
\usepackage{caption}
\usepackage{subcaption}
\usepackage{multirow}
\usepackage{amsthm}
\usepackage{algorithm}
\usepackage[noend]{algpseudocode}

\theoremstyle{definition}

\newlength\myindent
\makeatletter
\def\BState{\State\hskip-\ALG@thistlm}
\makeatother

\ifCLASSINFOpdf
\else
\fi
\begin{document}
\title{Communication Security in the Internet of Vehicles based Industrial Value Chain}
\author{
\IEEEauthorblockN{Yuanfang Chen\IEEEauthorrefmark{1}, Muhammad Alam\IEEEauthorrefmark{1}, Xiaohua Xu\IEEEauthorrefmark{2}}\\
\IEEEauthorblockA{
\IEEEauthorrefmark{1}School of Cyberspace, Hangzhou Dianzi University, China\\
\IEEEauthorrefmark{2}Kennesaw State University, USA}
}
\maketitle


\begin{abstract}
The Internet of Vehicles (IoV) is formed by connecting vehicles to Internet of Things.  It enables vehicles to ubiquitously access to the information of customers (drivers), suppliers, and even producers to structure an IoV-based industrial value chain.  Nevertheless, with the increase in vehicle networking and the information exchange among drivers, suppliers, and producers, communication security is emerging as a serious issue.  In this article, we provide an overview of the communication security, and a summary of security requirements.  Moreover, we clarify how security solutions can be used to conquer security challenges in the value chain.  Finally, an example of attack detection and identification in Electronic Control Units (ECUs) of vehicles is used to concretely illustrate the challenges.
\end{abstract}

\IEEEpeerreviewmaketitle


\section{Communication Security in the IoV-based Industrial Value Chain}
\label{sec:communication_security}
\begin{figure*}
  \centering
  \includegraphics[width=7in]{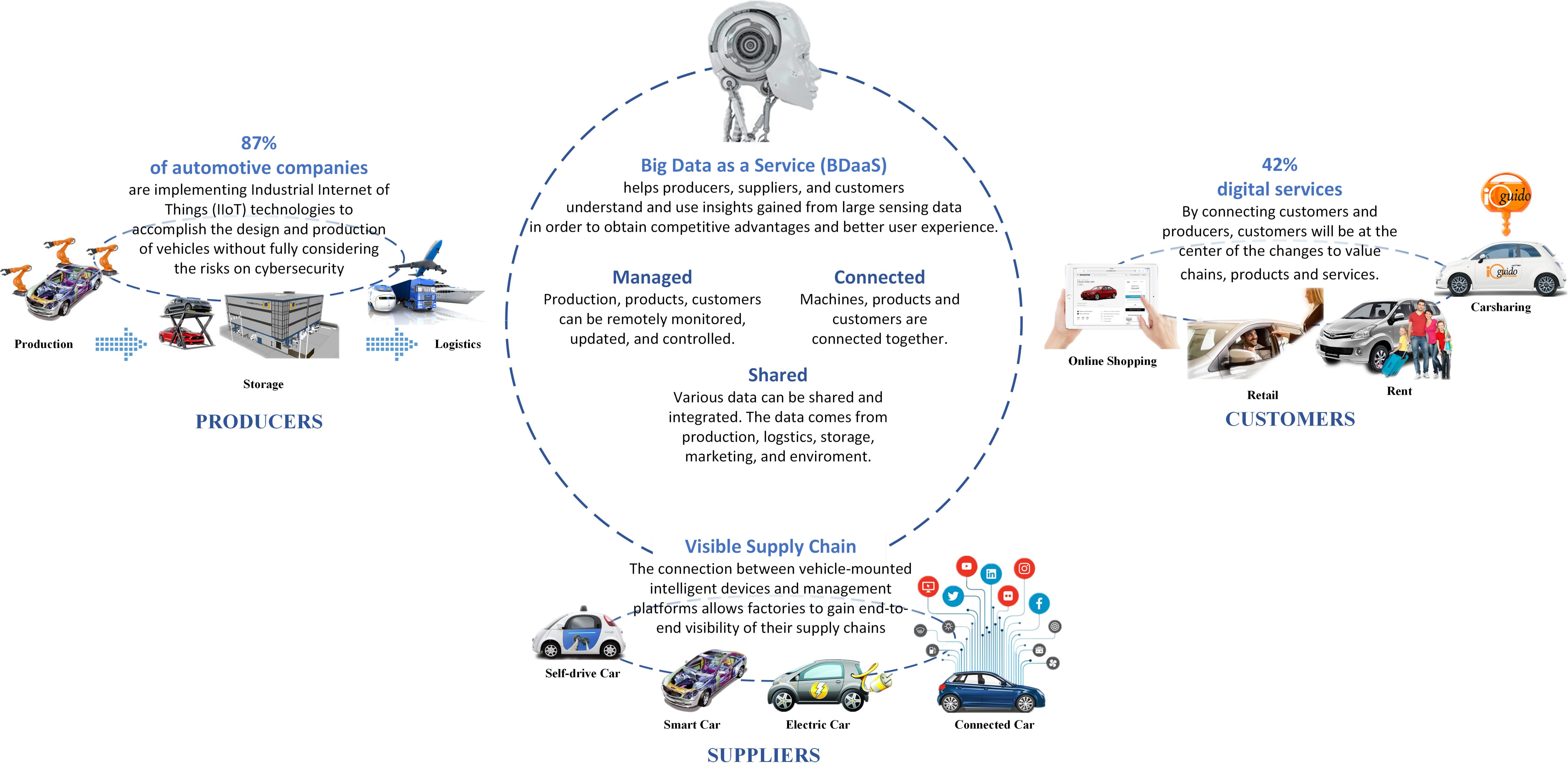}
  \caption{The industrial value chain of vehicles.}
  \label{fig:scenario}
\end{figure*}

The Internet of Vehicles (IoV) integrates different technologies, services, and standards to enable vehicles to ubiquitously access to the information of drivers (customers), suppliers, and even producers~\cite{contreras2018internet}.  As shown in Fig.~\ref{fig:scenario}, these three information sources are linked together by the value chain of the vehicle industry.  However, the connectivity of this industrial value chain increases the need for communication security~\cite{wollschlaeger2017future, da2014internet}.

Moreover, compared to the traditional vehicle industry, the advanced communication and IoV technologies have great help for vehicle industry to improve production efficiency.  However, security threats also spread to vehicle industry networks from information systems.  A variety of attacks can disturb the normal productive process of the vehicle industry~\cite{barry2018policy}.

The last two years, the infrastructure for the communications of the IoV has been frequently attacked~\cite{liu2017vehicle}.  In addition, the devices in the industrial value chain generate and exchange huge security, safety-critical and privacy-sensitive data, which is appealing to attackers~\cite{ji2018investigating,checkoway2011comprehensive}.  In~\cite{zhang2015securing}, Zhang has stated the vulnerabilities of communication security for the connected vehicles in detail.  For example, (i) the vulnerabilities of vehicle-to-networks infrastructure/cloud communications; (ii) the vulnerabilities of onboard communication networks; (iii) the vulnerabilities of vehicle-to-vehicle communications.  In the light of the above, the operating environment of devices in the value chain is vulnerable and unprotected, and is with serious communication security problems.  For example, the MP3 files which are under attacks infect a whole network of a vehicle quickly by taking advantage of the vulnerabilities of onboard communication networks~\cite{eiza2017driving}.  Once adversaries get the control of vehicular communication systems, they can manoeuvre the functional modules of vehicles, such as brakes, accelerators, and steering.  A demonstration has shown how attackers control a Jeep Cherokee by a vehicular communication system, while the Jeep is on the move~\cite{cho2016fingerprinting}, in the Black Hat cybersecurity conference of 2018.


\section{Security Requirements}
\label{sec:requirements}
There are several requirements that need to be followed, when we develop methods to achieve the communication security in the IoV-based industrial value chain~\cite{hussain2018secure}. Table~\ref{tab:table} summarize some of requirements for the communication security.
\begin{table*}[!ht]
\centering
\caption{Security requirements of the communications in the IoV-based industrial value chain}
\label{tab:table}
\vspace{0.2mm}
\renewcommand{\arraystretch}{2}
\begin{tabular}{llp{8cm}}
\bottomrule
\rowcolor{black!15}Security requirement & The involved parts in the value chain & Description\\
\bottomrule
\cellcolor{black!5}Authentication & \cellcolor{black!5}Producers, suppliers, and customers & \cellcolor{black!5}Verifying the identities of devices which transfer data in the value chain.\\
Integrity & Producers, suppliers, and customers & Checking whether the transmitted and received data have been delivered correctly.\\
\cellcolor{black!5}Confidentiality & \cellcolor{black!5}Producers, suppliers, and customers & \cellcolor{black!5}Protecting the confidential data transmission which occurs among the different devices participating in the value chain.\\
Access control & Producers, suppliers, and customers & Controlling the access permission of the available services for the devices in the value chain.\\
\cellcolor{black!5}Non-repudiation & \cellcolor{black!5}Producers, suppliers, and customers & \cellcolor{black!5}Ensuring that a device cannot deny the authenticity of another device in the value chain.\\
Availability & Producers, suppliers, and customers & Ensuring the communications among devices of the value chain in different conditions.\\
\cellcolor{black!5}Anti-interference & \cellcolor{black!5}Producers, suppliers, and customers & \cellcolor{black!5}Preventing malicious devices which send interfering messages to interrupt the communications among devices in the value chain.\\
\bottomrule	
\end{tabular}
\end{table*}

From Table~\ref{tab:table}, we notice that these seven security requirements are critical to all three parts of the value chain, producers, suppliers, and customers.  Moreover, there are many wireless devices in the value chain, so the requirement, anti-jamming, is proposed to prevent the interference from malicious devices.


\section{Solutions for Security Challenges}
\label{sec:solutions_challenges}
In the IoV-based industrial value chain, the challenges for communication security are mainly from two parts: industrial communication systems of vehicles, and vehicular communication systems, and there are three particular research challenges that cannot be handled by existing security strategies alone:
\begin{itemize}
  \item Data-sharing trust.  In the industrial communication systems of vehicles and vehicular communication systems, the trustworthiness of messages sent by relevant devices is primarily determined by the trustworthiness of these communicated devices.  Therefore, it is necessary to ensure the trustworthiness of these devices.  Nevertheless, in the dynamic communication environment, to establish the trustworthiness, one device is unable to achieve this alone, because the data which are used in the trust establishment come from multiple different devices.  On this basis, cooperative management is needed for the data-sharing trust.
  \item Attack detection.  (i) It is hard to confirm the cause of data anomaly.  Industrial communication systems of vehicles and vehicular communication systems are highly volatile in data transmission, so it is easy to make data transmission errors occur.  In such systems, if we want to conduct attack detection, it is a challenge to confirm the cause of data anomaly: does the attack or does the data transmission error cause the data anomaly?  (ii) it is low in detection efficiency.  This issue is generally considered as arising from the lack of studies on the nature of the attacks from communication systems.  It calls for the exploration and development of accurate processing schemes with the deep analysis for the features of the attacks from the communication systems; (iii) it is low in throughput, and high in costs.  Distributed detection paradigms are widely used in the attack detection, so the high data transfer rate and throughput are required to ensure the distributed communications.  Moreover, there are many devices in the value chain, which are limited in communication capability.  For this reason, the cost effectiveness of communications is important to the attack detection; (iv) it lacks in appropriate metrics and assessment methodologies, as well as a general framework to evaluate and compare alternative attack detection methods in the industrial communication systems of vehicles and vehicular communication systems.
  \item Attack location identification.  It is important to identify the attack locations in the wireless and mobile communication environment.  For example, in the vehicular communication system of a vehicle, no matter how well a method detects attacks, if the method does not get which Electronic Control Units (ECUs) are attacked, and further does not know which ECUs need to be isolated/patched, the vehicle remains insecure and unsafe.  From the cost aspect, if we can isolate/patch the ECUs which are under attacks, rather than blindly treat all ECUs as being under the attacks, it will be more economical.
\end{itemize}

We provide the corresponding feasible solutions to the above mentioned security challenges.
\begin{itemize}
  \item Data trustworthiness should be partially attributed to the trustworthiness of communication devices.  The constructing of the trust among devices does not rely on a single source of data. In addition, deriving the trust in data is by collaborating multiple pieces of evidence from different devices.  It is a feasible method by weighing each individual piece of evidence, and meanwhile considering various trust metrics~\cite{lu2018survey}.  Data/relevant communication devices and their respective weights input into the decision logic which is used to output the trustworthiness level of these data/devices.  In addition, Bayesian inference and dempster-shafer theory are available methods to derive the trustworthiness of the data/devices.  Bayesian inference puts prior knowledge into account, and the dempster-shafer theory provides the uncertainty of the data/devices.  The uncertainty can be considered as a kind of supporting or refuting evidence to make a more realistic decision for the trustworthiness level of the data/devices.
  \item In vehicle communication systems, attacks can destroy the function and performance of safety-critical communication components to threaten driving safety.  In~\cite{shi2017finite}, Dawei~\emph{et al.} have proposed that if attacks can be timely detected in a vehicle communication system, the damage for the overall driving system is able to be reduced to a tolerable range.  In this regard, attack detection is important and necessary to ensure the driving safety of vehicles.  Two kinds of schemes are feasible to defend against attacks in communication systems: (i) protecting the important system components beforehand; (ii) identifying the data errors and loss caused by attacks afterwards.  It is possible to achieve the first scheme by deploying attack defense modules in the vehicle production stage, or setting redundant components and communication pathways.  For example, Phasor Measurement Units (PMUs) deployment directly monitors state information to make the systems free from attacks.  For the second scheme, cryptography has been used against attacks. However, such type of approaches cannot provide adequate protection to defense adversaries.  Four alternative detection approaches have been proposed for the second scheme: (i) Bayesian detection; (ii) weighted least square detection; (iii) kalman filter based $\mathcal{X}^{2}$ detector; and (iv) feature based detection.  Different schemes are appropriate for different attacks and attack scenarios to detect the existence and the type of the attacks.
  \item The thus-derived technique can be used to identify attack locations~\cite{cho2016fingerprinting}, when attack messages are periodically injected.  However, in most attack scenarios, attack messages are not injected periodically.  It is necessary to design a method which doesn't need to consider when and how the attack messages are injected.  Fingerprinting methods have been proposed to fingerprint communication components, and then the fingerprints of components are used to identify attack locations.  For example, in vehicular communication systems, voltage measurements are special for each ECUs, and can be used to fingerprint the ECUs.  In~\cite{choi2018identifying}, the proposed identification method uses the voltage signal values which are corresponding to the extended identifier field of the Controller Area Networks (CANs) data frame, and the voltage signals are generated by CAN transceivers of ECUs.  However, this solution is unpractical for identifying attack locations: (i) modeling is done by using batch learning; it is infeasible to be used in the attack location identification which is a kind of real-time application; (ii) there are unpredictable changes in the CAN bus and adversary's behaviors, which can cause false identification.  In~\cite{cho2017viden}, Viden was proposed.  It provides adaptability through the real-time update for ECUs' fingerprints, and it doesn't restrict the CAN message's type, and the CAN bus's speed.
\end{itemize}


\section{An Example of Communication Security}
The vehicle industry has come to incorporate the latest information and communication technology, and increasingly replace vehicles' mechanical components with electronic components.  The ECUs of a vehicle can communicate with each other in a vehicle communication system, and communicate with other external information systems via the on-board diagnostics (OBD) interface, which makes the vehicle both safer and easier to drive.  The CAN bus is the current standard for in-vehicle communications.  Unfortunately, it does not currently offer protection against security attacks.  It does not allow for message authentication, and hence is open to attacks that replay ECU messages for malicious purposes.  The classic cryptographic method of the message authentication code (MAC) is not feasible, since the CAN data frame is not long enough to include a sufficiently long MAC to provide effective authentication.  To mitigate this, some researchers consider modeling the intervals of messages to detect and identify attacks.  We investigate the feasibility of this way, to be as a concrete demonstration example to illustrate the challenges of communication security in the IoV-based industrial value chain.  In this investigation, we measure and use the periodic intervals of the transmitted messages in vehicle communication systems to fingerprint ECUs.  Such fingerprinting can be used to identify attack locations and even attack types.

In our experiments, three types of attacks are considered: DoS attack, Fuzzy attack, and Impersonation attack.  Moreover, multiple in-vehicle ECUs can be remotely compromised by adversaries, and the three types of attacks are able to make full control to the compromised ECUs and access to memory data.  Meanwhile, the adversaries mount attacks by injecting arbitrary attack messages.

\textbf{DoS Attack.}  For this kind of attack, in our study, the attackers use the messages with the theoretically highest priority identifier 0x000 to occupy the CAN bus.  In this scenario, there are two injected 0x000 messages from the compromised ECU which is attacked by the DoS, and these two messages inevitably delay the reception of the response messages 0x2C0 and 0x5A2 from other two normal ECUs.  In a CAN bus of a vehicle, if a kind of message increases the occupancy for the bus, it will induce the latencies of other messages in this bus, and further may cause serious threats to the driving due to the unavailability of response messages to the commands from the driver.

\textbf{Fuzzy Attack.}  For this kind of attack, in our study, the messages with randomly spoofed identifiers are injected into a CAN bus by attackers.  Unlike the DoS attack which delays normal messages by occupying the CAN bus, the fuzzy attack makes the function of a vehicle paralyzed such as shaking the steering wheel, irregularly turning on and off signal lamps, and automatically changing the gear shift.  For example, the normal ECU A and ECU C send the messages 0x2C0 and 0x5A2 to the CAN bus, and meanwhile the compromised ECU B sends the spoofed 0x2C0 and 0x5A2 to the CAN bus, to make steering wheels shaky.

\textbf{Impersonation Attack.}  For this kind of attack, in our study, it can stop the message transmission of a target ECU, and plant/manipulate an impersonating ECU.  That is, once the target ECU stops to transmit messages in a CAN bus, all the messages which are sent from the ECU will be removed from the bus, and the messages sent from the impersonating ECU will make replacement.  For example, ECU B is an impersonating ECU, and it stops the message transmission of a target ECU A, the messages from ECU A are removed from the CAN bus, and the ECU B replaces the ECU A to send messages.

\subsection{Experimental Setup}
We use four CAN nodes to build a prototype.  Nodes A, B, C are normal, and node X is compromised by attackers.  These four nodes are linked together by a CAN bus, and connected to a data centre via wireless mode.  Each node supports the ISO15765 CAN bus protocol with 250Kbps and 500Kbps bus speed, and the 16PIN OBD interface with a wireless communication component.  This wireless interface is used to provide external communication capability.

In this prototype, the CAN bus speed is set as 500Kbps.  The node A is set to send the messages with ID=0x00000001, 0x00000002, and 0x00000003, the node B is to send the messages with ID=0x00000004, 0x00000005, and 0x00000006, the node C is to send the messages with ID=0x00000007, 0x00000008, and 0x00000009, and the period of transmission for these messages is 50ms.  The node X is set to be compromised by attackers, and it performs attacks to attack any target ECU.

We use the calculated clock skew from ECU messages to be as the fingerprint of the ECU.  In the CAN bus, it lacks clock synchronization, so the message sending of ECUs in the bus is considered to be unsynchronized, and the clock skew calculated from an unsynchronized ECU depends solely on its local clock, so it is distinct.  It can therefore be considered as the fingerprint of an ECU.

The slope of message sending's accumulated clock offset is the clock skew, and the accumulated clock offset is calculated with summing the average clock offsets\footnote{In many previous work, the clock offset of message sending is defined as the difference between the observed clock $C_{i}$ and the true clock $C_{true}$.} every certain message samples.  Various studies have exploited this fact to fingerprint physical devices.  However, unlike many previous approaches that exploit the difference between the observed clock $C_{i}$ and the true clock $C_{true}$, we exploit the periodicity of message sending.  For example, ECU A periodically sends a message to ECU B every $T$ms, but due to the clock skew, periodic messages are sent with small offsets from the ideal time points (e.g., $T$, $2T$, $3T$, ...).  That is, let $t=0$ be the time when the first message was sent from the ECU A, and let $O_{i}$ be the clock offset of the ECU A when it sends the $i^{th}$ message since $t=0$.  Then, after a network delay of $d_{i}$, the ECU B will receive the message from the ECU A, and put an arrival timestamp of $iT+O_{i}+d_{i}$.  Thus, the interval between each arrival timestamps is $T+\bigtriangleup O_{i}+\bigtriangleup d_{i}$, and the expected value of the intervals is $T+E[\bigtriangleup O_{i}+\bigtriangleup d_{i}]$.

On this basis, we can work out the estimated arrival time of the $i^{th}$ message $i(T+E[\bigtriangleup O_{i}+\bigtriangleup d_{i}])+d_{0}$, where $d_{0}$ is the arrival timestamp of the first message, whereas the actual measured arrival time is $iT+O_{i}+d_{i}$.  The average difference between the estimated and the measured is $E[i(E[\bigtriangleup O_{i}+\bigtriangleup d_{i}])+O_{i}+\bigtriangleup d] \approx E[O_{i}]$.  It means that the average clock offset $E[O_{i}]$ can be estimated.

In summary, the clock skew of message sending can be used to fingerprint an ECU.

\subsection{Experimental Results and Analysis}
\begin{figure*}[!ht]
  \centering
  \includegraphics[width=7.6in]{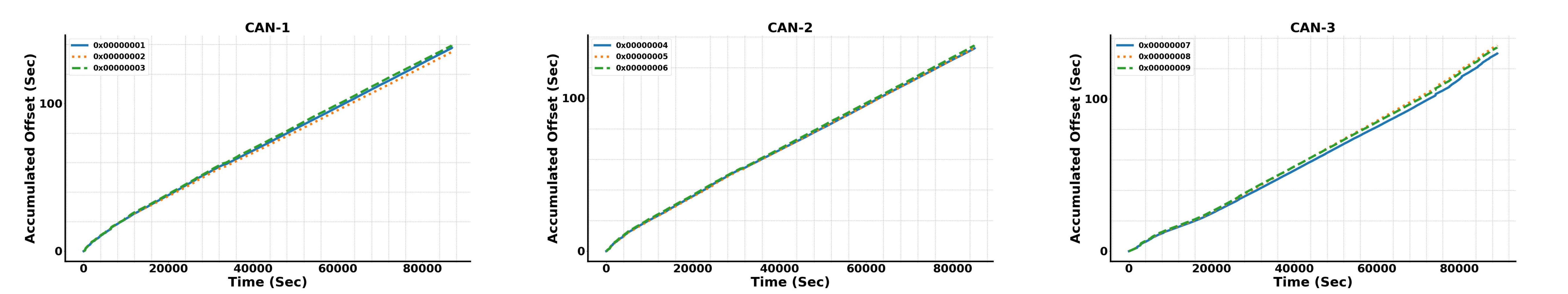}
  \caption{Accumulated clock offsets of messages 0x00000001-0x00000009 for three normal ECUs.}
  \label{fig:oacc_normal}
\end{figure*}

\begin{figure*}[!ht]
  \centering
  \includegraphics[width=7.6in]{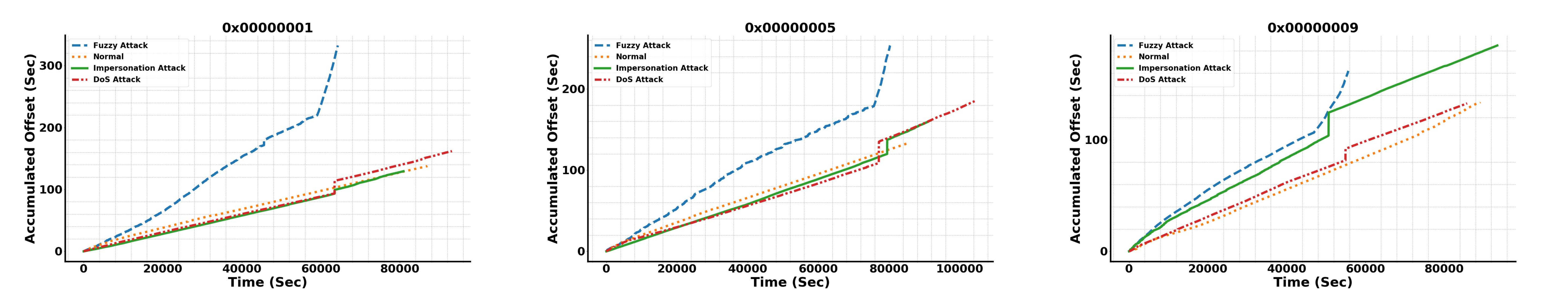}
  \caption{Accumulated clock offsets in different attack scenarios.}
  \label{fig:oacc_three_attacks}
\end{figure*}

The validity of fingerprinting ECUs is evaluated by estimating and observing clock skews, and the evaluation result is shown in Fig.~\ref{fig:oacc_normal}.  Then, we launch DoS, fuzzy and impersonation attacks.  The effectiveness of fingerprinting in detecting and pinpointing attacks is evaluated, and the result is shown in Fig.~\ref{fig:oacc_three_attacks}.

Figure~\ref{fig:oacc_normal} gives the accumulated clock offsets of messages 0x00000001-0x00000009 for three normal ECUs, and the slopes of the curves are the clock skews of messages.  Note that all of these accumulated clock offsets are linear in time and with tiny fluctuation.  That is, each clock skew is constant, messages 0x00000001-0x00000003, 0x00000004-0x00000006, and 0x00000007-0x00000009 are sent from the node A, B and C respectively, and they have respective constant clock skews.  Moreover, there is tiny fluctuation in each clock skew.  The fluctuation roots in this part in calculating the average clock offset: $E[\bigtriangleup d]$, the difference between $d_{i}$ and $d_{0}$.  However, in an actual CAN bus, it has a high transmission speed, so the transmission delay is very small, and the difference between $d_{i}$ and $d_{0}$ is tiny.  Therefore, the ECU's clock skew is able to be used to fingerprint the ECU.

For the evaluation of defending against DoS, impersonation and fuzzy attacks, the node X is programmed to inject the attack messages with ID=0x00000001-0x00000009 into three ECUs respectively.

Figure~\ref{fig:oacc_three_attacks} shows how accumulated clock offsets change for messages 0x00000001 (sent from the node A), 0x00000005 (sent from the node B) and 0x00000009 (sent from the node C) in the presence and absence of three kinds of attacks on the CAN bus prototype.  From observing these results, there are abrupt changes in the slopes at the attack points.  For the DoS attack, the attack messages are sent at the attack point to occupy the CAN bus, so excepting where the attack occurred, the message sending is still periodic on the CAN bus.  For the impersonation attack, the attack messages replace the normal messages to send on the CAN bus, and as the DoS attack, the message sending is periodic as well on the CAN bus.  The fuzzy attack injects attack messages with the same ID as the sending messages on the CAN bus, rather than occupying the CAN bus.  It means that there are two messages with the same ID on the CAN bus, which makes the message sending aperiodic, so the slope is changed after the attack point.

\subsection{Discussion}
In this section, we discuss the limitation of fingerprinting ECUs.

The limitation of fingerprinting ECUs is that it can only extract clock skews from periodic messages; it would be difficult to fingerprint ECUs which are sending aperiodic messages.  Unfortunately, message sending is not always periodic in vehicle communication systems.  Moreover, almost all of the message injection for an attacker is aperiodic, although it is still able to detect attacks by fingerprinting, it would not be able to pinpoint where the attack message came from, i.e., finding the ECU which is compromised by the attacks that launch the aperiodic messages.  For this limitation, we would like to find new features other than clock skews, which can be used to fingerprint ECUs, regardless of whether attacks/ECUs send messages periodically or aperiodically.


\section{Conclusions}
\label{sec:Conclusion}
Vehicle industry has come to incorporate the latest communications technology to combine the information from drivers, suppliers, and even producers.  It makes the vehicle both safer and easier to drive.  However, current communication systems for the vehicle industry are not sufficiently enhanced to bear security risks.  Security attacks against the communication systems may threaten the physical safety of drivers.

To protect communication security in the industrial value chain of vehicles with a large number of heterogeneous devices, we need a holistic cybersecurity framework to be compatible with all heterogeneous devices and communication protocols.  This framework needs to cover all abstraction layers of these heterogeneous devices, and to be across the boundaries of the communications among these devices.  However, existing security solutions are inappropriate for the heterogeneous environment with capability-limited devices.  Moreover, they cannot meet real-time requirements to defense attacks as well.  It is necessary to design and develop appropriate security methods for the heterogeneous environment, including new primitives that are universal to heterogeneous and capability-limited devices, and are resilient to instant attacks.

In this article, we indicate the requirements for communication security, and clarify how security solutions can be used to conquer security challenges in the value chain.  Moreover, as an example, we verify the feasibility of fingerprinting ECUs as the universal primitives to detect and identify the attacks in heterogeneous environments.


\section*{Acknowledgments}
This work was supported by the National Natural Science Foundation of China (Grant No. 61802097), and the Project of Qianjiang Talent (Grant No. QJD1802020).


\bibliographystyle{IEEEtran}
\bibliography{IEEEfull}


\end{document}